\documentstyle [12pt,epsf] {article}
\renewcommand{\baselinestretch}{1.2}

\begin{document}
\def\be{\begin{eqnarray}}
\def\en{\end{eqnarray}}
\def\non{\nonumber}
\def\xx{{\bf XX~}}
\def\la{\langle}
\def\ra{\rangle}
\def\bt#1{\hbox{$\lower5pt\hbox{$#1$}\atop\sim$}}
\def\sqr#1#2{{\vcenter{\hrule height.#2pt
        \hbox{\vrule width.#2pt height #1pt \kern#1pt
        \vrule width.#2pt}
        \hrule height.#2pt}}}
\def\ri{\rightarrow}
\def\pc{{\rm pc}}
\def\pv{{\rm pv}}
\def\up{\uparrow}
\def\dw{\downarrow}
\def\halfm{{1\over 2}^{^-}}
\def\halfp{{1\over 2}^{^+}}
\def\half{{1\over 2}}
\def\thalf{{3\over 2}}
\def\lamc{{\Lambda_c^{^+}}}
\def\xia{\Xi^{+A}_c}
\def\xi0a{\Xi^{0A}_c}
\def\siga{\Sigma^{^+}(8,P_{1/2})_a}
\def\sigc{\Sigma^{^+}(8,P_{3/2})}
\def\sigz{\Sigma^{^0}}
\def\tx{\tilde{X}}
\def\uup{u^{\uparrow}}
\def\udw{u^{\downarrow}}
\def\dup{d^{\uparrow}}
\def\ddw{d^{\downarrow}}
\def\sup{s^{\uparrow}}
\def\sdw{s^{\downarrow}}
\def\cup{c^\uparrow}
\def\cdw{c^\downarrow}
\def\tuup{\tilde{u}^{\uparrow}}
\def\tudw{\tilde{u}^{\downarrow}}
\def\tdup{\tilde{d}^{\uparrow}}
\def\tddw{\tilde{d}^{\downarrow}}
\def\tsup{\tilde{s}^{\uparrow}}
\def\tsdw{\tilde{s}^{\downarrow}}
\def\buup{\bt{u}^\up}
\def\budw{\bt{u}^\dw}
\def\bdup{\bt{d}^\up}
\def\bddw{\bt{d}^\dw}
\def\bsup{\bt{s}^\up}
\def\bsdw{\bt{s}^\dw}
\def\buupl{\bt{u}^{^\big{\uparrow}}}
\def\bdupl{\bt{d}^{^\big{\uparrow}}}
\def\bsupl{\bt{s}^{^\big{\uparrow}}}
\def\aa{{\cal A }_A}
\def\as{{\cal A }_S}
\def\bsp{{\cal B }'_S}
\def\bap{{\cal B }'_A}
\def\ba{{\cal B }_A}
\def\cap{{\cal C }'_A}
\def\coa{{\cal C }_{1A}}
\def\cta{{\cal C }_{2A}}
\def\csp{{\cal C }'_S}
\def\cts{{\cal C }_{2S}}
\def\ea{{\cal E }_A}
\def\es{{\cal E }_S}
\def\pa{{\cal E }'_A}
\def\ps{{\cal E }'_S}
\def\a{{\cal A}}
\def\b{{\cal B}}
\def\c{{\cal C}}
\def\d{{\cal D}}
\def\e{{\cal E}}
\def\ra{\rangle}
\def\la{\langle}
\def\jo{\journal}
\def\tr{{\rm Tr}}
\def\et{{\it et al.}}

\def\pr{{\sl Phys. Rev.}~}
\def\prl{{\sl Phys. Rev. Lett.}~}
\def\pl{{\sl Phys. Lett.}~}
\def\np{{\sl Nucl. Phys.}~}
\def\zp{{\sl Z. Phys.}~}

\def\half{{1\over 2}}
\def\thalf{{3\over 2}}
\def\lamc{{\Lambda_c^{^+}}}
\def\xia{\Xi^{+A}_c}
\def\xi0a{\Xi^{0A}_c}
\def\siga{\Sigma^{^+}(8,P_{1/2})_a}
\def\sigc{\Sigma^{^+}(8,P_{3/2})}
\def\sigz{\Sigma^{^0}}
\def\tx{\tilde{X}}
\def\uup{u^{\uparrow}}
\def\udw{u^{\downarrow}}
\def\dup{d^{\uparrow}}
\def\ddw{d^{\downarrow}}
\def\sup{s^{\uparrow}}
\def\sdw{s^{\downarrow}}
\def\cup{c^\uparrow}
\def\cdw{c^\downarrow}
\def\tuup{\tilde{u}^{\uparrow}}
\def\tudw{\tilde{u}^{\downarrow}}
\def\tdup{\tilde{d}^{\uparrow}}
\def\tddw{\tilde{d}^{\downarrow}}
\def\tsup{\tilde{s}^{\uparrow}}
\def\tsdw{\tilde{s}^{\downarrow}}
\def\buup{\bt{u}^\up}
\def\budw{\bt{u}^\dw}
\def\bdup{\bt{d}^\up}
\def\bddw{\bt{d}^\dw}
\def\bsup{\bt{s}^\up}
\def\bsdw{\bt{s}^\dw}
\def\buupl{\bt{u}^{^\big{\uparrow}}}
\def\bdupl{\bt{d}^{^\big{\uparrow}}}
\def\bsupl{\bt{s}^{^\big{\uparrow}}}
\def\aa{{\cal A }_A}
\def\as{{\cal A }_S}
\def\bsp{{\cal B }'_S}
\def\bap{{\cal B }'_A}
\def\ba{{\cal B }_A}
\def\cap{{\cal C }'_A}
\def\coa{{\cal C }_{1A}}
\def\cta{{\cal C }_{2A}}
\def\csp{{\cal C }'_S}
\def\cts{{\cal C }_{2S}}
\def\ea{{\cal E }_A}
\def\es{{\cal E }_S}
\def\pa{{\cal E }'_A}
\def\ps{{\cal E }'_S}
\def\a{{\cal A}}
\def\b{{\cal B}}
\def\c{{\cal C}}
\def\d{{\cal D}}
\def\e{{\cal E}}
\def\ra{\rangle}
\def\la{\langle}
\def\jo{\journal}
\def\tr{{\rm Tr}}
\def\et{{\it et al.}}

\def\pr{{\sl Phys. Rev.}~}
\def\prl{{\sl Phys. Rev. Lett.}~}
\def\pl{{\sl Phys. Lett.}~}
\def\np{{\sl Nucl. Phys.}~}
\def\zp{{\sl Z. Phys.}~}

\font\el=cmbx10 scaled \magstep2
{\obeylines
\hfill UCDPHY-PUB-95-26
\hfill IP-ASTP-13-95
\hfill January, 1996}

\vskip 1.5 cm

\centerline{\large\bf Analysis of Two-Body Decays of Charmed Baryons }
\centerline{\large\bf Using the Quark-Diagram Scheme}
\medskip
\bigskip
\medskip
\centerline{\bf Ling-Lie Chau}
\medskip
\centerline{ Physics Department, University of California at Davis,
California 95616}
\bigskip
\centerline{\bf Hai-Yang Cheng and B. Tseng}
\medskip
\centerline{Institute of Physics, Academia Sinica, Taipei, 
Taiwan 11529}
\bigskip
\bigskip
\bigskip
\centerline{\bf Abstract}
\bigskip
  
We give a general formulation of the quark-diagram scheme for  the
nonleptonic weak decays of baryons. We apply it to all the decays  
of the antitriplet and  sextet  charmed baryons and
express their decay amplitudes in terms of the quark-diagram amplitudes.  We
have also given parametrizations for the effects of final-state 
interactions.  For SU(3) violation effects, we only
parametrize those in the horizontal $W$-loop quark diagrams whose 
contributions are
solely due to SU(3)-violation effects. In the absence of all these effects,
there are many  relations among various decay modes. Some of
the relations are valid even in the presence of final-state interactions when
each decay amplitude in the  relation contains only a single phase shift.
All these relations provide  useful frameworks 
to compare with future experiments and to find out the effects of final-state
interactions and SU(3) symmetry violations.

\pagebreak
%
%
\noindent{\bf I. Introduction}

  The study of charmed baryon physics is of current interest [1]. Many
nonleptonic weak decay modes of the charmed baryons $\Lambda_c^+$,
$\Xi_c^{0A}$ and $\Xi_c^{+A}$ have been measured [2] and more data
are expected in the near future. Apart from model calculations [3-5], it is 
useful to study the nonleptonic weak decays
in a way which is as model independent as possible. The two-body nonleptonic
decays of charmed baryons have been analyzed in terms of
SU(3)-irreducible-representation
(SU(3)-IR)  amplitudes [6,7]. However, the quark-diagram
scheme (i.e., analyzing the decays in terms of quark-diagram amplitudes) 
 has the advantage that it is more intuitive and easier for implementing 
model calculations.
It has been successfully applied to the hadronic weak decays of
charmed and bottom mesons [8,9]. It has provided a
framework with which we not only can do the least-model-dependent data
analysis and give predictions
but also make evaluations of theoretical model 
calculations. In this paper we give a
general and unified formulation (always using the orthonormal 
quark states and independent of SU(6) symmetry) of the quark-diagram scheme
for the
nonleptonic weak decays of baryons, which can be useful for all baryon 
(charm and bottom) nonleptonic decays.
Here we apply it to all the two-body hadronic decays (quark-mixing allowed,
suppressed,
and doubly-suppressed) of the antitriplet and  sextet  charmed baryons and
express them in terms of
the quark-diagram amplitudes.  We find consistent comparisons with the
SU(3)-IR results of Ref.~[6].
In addition,  the symmetry properties of the initial and final baryon
wave functions in conjunction with the Pati-Woo theorem [11] for weak decays
enable us to obtain more specific results than those from the SU(3)-IR
scheme. 

 We have also given parametrizations for effects of final-state 
interactions. For SU(3) violations we indicate only those  arising in the
horizontal $W$-loop quark diagrams whose contributions are solely due to
SU(3)
violation effects.  (See more detailed discussion about this at the end of  
Section III.a.) In the absence of these effects, there are
many relations among various decay modes.  Some of
the relations are valid even in the presence of final-state interactions when
each decay amplitude in the  relation contains only a single phase shift.
It will be interesting to  compare  these relations with future experimental
data. They provide a useful framework to find out the effects of final-state
interactions and SU(3) violations.

Earlier Kohara had given a quark-diagram formulation for the
quark-mixing-allowed decays of the antitriplet charmed baryons [10], with
which our results agree.  However, for the decay products
containing an octet baryon he used
a basis of quark states which is not orthonormal (which is all right though
more complicated) and its
choice seems to be based upon exact  SU(6) symmetry (which is not all right
since SU(6) is not an
exact symmetry). The fact that our formulation  demonstrates
that the quark diagram results are independent of whether one uses exact
SU(6) or not helps to
clarify this point for his formulation and explain why our results can
possibly agree. (For detailed
comments see Sections II after Eq.~(28) and for exact relations between
amplitudes see Eq.~(63) in Section  IV.b.)

   In the framework of the quark-diagram scheme, all nonleptonic meson decays
can be expressed  in terms of six quark-diagram amplitudes [8]:
${\cal A}$, the external ${\it W}$-emission diagram; ${\cal B}$, the internal
${\it W}$-emission diagram; ${\cal C}$, the ${\it W}$-exchange diagram;
 ${\cal D}$, the ${\it W}$-annihilation diagram; ${\cal E}$, the horizontal
${\it  W}$-loop diagram; and ${\cal F}$, the vertical  ${\it W}$-loop
diagram. These quark diagrams are specific and well-defined physical
quantities. They are classified according to the topology of first-order
weak interactions, with all QCD strong-interaction effects included.
It is important to emphasize that strong interactions do not alter the
classification of these diagrams.
These quark diagrams have a one-to-one
correspondence to those amplitudes classified according to SU(3)
irreducible representations.

For the baryon decays, we can easily show by diagram drawing that the  
${\cal D}$ and ${\cal F}$ types of amplitudes do not contribute. However,
there are more
possibilities in drawing the ${\cal C}$ and ${\cal E}$ types of amplitudes.
More importantly, baryons
being made out of three quarks, in contrast to two quarks for the mesons,
bring along many essential
complications. Though  many textbooks [12] have discussed the baryon wave
functions, we need to
carefully develop the proper formulation suitable for the construction of
the quark
diagram scheme for the  baryon decays. This is what we discuss in Section II,
where the relations
between the quark states and the baryon states  are derived. We then apply
this general results to
the specific decays of the charmed baryons. In Sections III and IV we give
the quark-diagram
formulation for the two-body decays of  antitriplet charmed baryons into a
pseudoscalar meson and a
baryon (decuplet and octet); first for the case without effects of
final-state interactions and 
SU(3) violations, and then for the case  with these
effects. We discuss their experimental implications and comment on previous
related theoretical
work. Section V is devoted to studying the  nonleptonic weak decays of sextet
charmed
baryons. In Section VI
we give a few concluding remarks.

\noindent {\bf II. Quark States and Particle States}

To develop a quark diagram scheme we need to fully understand the
relation between the quark states and the particle states.  Baryons
are made out of three ${1\over 2}$-spin quarks.  The baryon states form
irreducible representations
of SU(3)-flavor and SU(2)-spin from the tensor-product states of flavor and
spin of
three quarks which are written as the following orthonormalized states:
\be
|q_1,S_{1z};~q_2,S_{2z};~q_3,S_{3z}\rangle = |q_1~q_2~q_3\rangle
|S_{1z}~S_{2z}~S_{3z}\rangle~.
\en
There are $3\times 3\times 3$ = 27 flavor states $|q_1~q_2~q_3\rangle$ and 
$2\times 2\times 2$ = 8 spin states $|S_{1z}~S_{2z}~S_{3z}\rangle$.

Let us first discuss the flavor irreducible representation states of the
three
quarks.  The 27 tri-quark states can be decomposed into 
$[8]_A,~[8]_S,~[1]_A$, and $[10]_{S_t}$ irreducible representations, 
denoted by the orthonormalized states:
\be
|\psi(8)_A\rangle,~ |\psi(8)_S\rangle,~ |\psi(1)_A
\rangle,{\rm ~~and~~}|\psi(10)_{S_t}\rangle~.
\en
The transformation between the two bases, Eq.~(1) and (2), can be written 
in a $27\times 27$ matrix which is block-diagonalized into the following 
sub-matrix transformations:
\be
\left(\matrix{|\psi^k(8)_S\rangle\cr
|\psi^k(8)_A\rangle\cr
|\psi^k(10)_{S_t}\rangle\cr}\right)\quad = \quad \left(
\matrix{{1\over\sqrt 6}&{1\over\sqrt 6}& -{2\over\sqrt 6}\cr
{1\over\sqrt 2}& -{1\over\sqrt 2}&0\cr
{1\over\sqrt 3}&{1\over\sqrt 3}&{1\over\sqrt 3}\cr}\right)\quad\left(
\matrix{|q_a~q_b~q_a\rangle\cr
|q_b~q_a~q_a\rangle\cr
|q_a~q_a~q_b\rangle\cr}\right)~,
\en
where $k$ can be the proton, neutron, $\Sigma^+,~\Sigma^-,~\Xi^0,~\Xi^-$ 
types and all of which have two identical quarks.  
There are 6 of such $3\times 3$ matrix equations totalling the
transformations of the 18 states out of the 27.
Note that the subscripts $A$ and $S$ signify the antisymmetry and 
symmetry, respectively, between the first two
 quarks; the subscript $S_t$ denotes the total symmetry among the
 three quarks.  Then there are the following transformations of the 6 
 states with all three quarks being different:
\be
\left( \matrix{|\psi^\Sigma(8)_S\rangle\cr
|\psi^\Sigma(8)_A\rangle\cr
|\psi^\Lambda(8)_S\rangle\cr
|\psi^\Lambda(8)_A\rangle\cr
|\psi^{\Lambda_1}(1)_A\rangle\cr
|\psi^\Sigma(10)_{S_t}\rangle\cr}\right) \quad =\quad
\left(\matrix{
{1\over\sqrt{12}}&{1\over\sqrt{12}}&{1\over\sqrt{
12}}&{1\over\sqrt{12}}&{-2\over\sqrt{12}}& {-2\over\sqrt{12}}\cr
{-1\over 2}&{1\over 2}& {-1\over 2}&{1\over 2}&0&0\cr 
{1\over 2}&{1\over 2}& {-1\over 2}& {-1\over 2}&0&0\cr 
{1\over\sqrt{12}}& {-1\over\sqrt{12}}& {-1\over\sqrt{
12}}&{1\over\sqrt{12}}& {-2\over\sqrt{12}}&{2\over\sqrt{12}}\cr
{1\over\sqrt 6}& {-1\over\sqrt 6}&{1\over\sqrt 6}& {-1\over\sqrt
6}&{1\over\sqrt 6}& {-1\over\sqrt 6}\cr
{1\over\sqrt 6}&{1\over\sqrt 6}&{1\over\sqrt 6}&{1\over\sqrt
6}&{1\over\sqrt 6}&{1\over\sqrt 6}\cr}\right)\quad \left(
\matrix{|sdu\rangle&\cr |dsu\rangle&\cr |sud\rangle&\cr
|usd\rangle&\cr |dus\rangle&\cr |uds\rangle&\cr}\right).
\en
Finally, there are the three states with all three identical quarks:
\be
|\Delta^{++}\rangle &=& |uuu\rangle~,   \\
|\Delta^-\rangle &=& |ddd\rangle~,    \\
|\Omega^-\rangle &=& |sss\rangle~.  
\en
They give three diagonal transformations. 
These 27 equations, Eqs.~(3) to (7), are actually equivalent to the following
27 equations:
\be
|\psi^k(8)_A\rangle &=& \sum_{q_i=u,s,d}~ |q_1~q_2~q_3\rangle~\langle 
q_1~q_2~q_3~|\psi^k(8)_A\rangle~,    \\
|\psi^k(8)_S\rangle &=& \sum_{q_i=u,s,d} ~|q_1~q_2~q_3\rangle~\langle 
q_1~q_2~q_3~|\psi^k(8)_S\rangle~,   \\
|\psi^k(1)_A\rangle &=& \sum_{q_i=u,s,d} ~|q_1~q_2~q_3\rangle~\langle 
q_1~q_2~q_3~|\psi^k(1)_A\rangle~,  \\
|\psi^k(10)_{S_t}\rangle &=& \sum_{q_i=u,s,d}~ |q_1~q_2~q_3\rangle~\langle 
q_1~q_2~q_3~|\psi^k(10)_{S_t}\rangle~,  
\en
where the superscript $k$ stands for the particles in the multiplets,
respectively.
These equations are obtained simply by multiplying the left-hand side
(l.h.s.) of these equations by the identity operator 
\be
\hat I=\sum_{q_i=u,d,s}~ |q_1~q_2~q_3\rangle~\langle 
q_1~q_2~q_3|~,
\en
which is the completeness of the orthonormal  $|q_1~q_2~q_3\rangle$-basis in
the 
tri-quark vector space. 
$\langle  q_1~q_2~q_3|\psi^k(\cdots)\rangle$ numbers in Eqs.~(8) to (11) are
precisely
those matrix elements in Eqs.~(3) to (7). 

Since the transformations, Eqs.~(3) to (7), are between two sets of 
orthonormal bases, the transformation matrix formed by these elements 
$\langle 
q_1~q_2~q_3|\psi^k(\cdots)\rangle$ is orthogonal. We can easily take the  
inverse of the
transformation (i.e. taking the transpose of the matrix) and express  the
quark states in terms
of the irreducible representation states, i.e., the particle states.  If the
basis vectors
are not orthonormal, not only such inverse will be harder to find but also
the completeness equation
will be   more complicated than that given by Eq.~(12).
Therefore, whenever possible, it is better to use the orthonormal basis.

Alternatively, we can also use the basis composed of the quark states that
are symmetric and antisymmetric in the first two quarks, i.e.,
\be
|\{q_a~q_b\}q_c\ra &\equiv &{1\over \sqrt{2}(1-\delta_{ab})+2\delta_{ab}}\,
\Big(|q_aq_bq_c\ra+|q_bq_aq_c\ra\Big), \\
|[q_a~q_b]q_c\ra &\equiv&
{1\over\sqrt{2}}\Big(|q_aq_bq_c\ra-|q_bq_aq_c\ra\Big
),   
\en
or inversely,
\be
|q_aq_bq_c\ra=\,{\sqrt{2}(1-\delta_{ab})+2\delta_{ab}\over 2}\,\Big(|\{q_a~
q_b\}q_c\ra+|[q_a~q_b]q_c\ra\Big). \en
In this basis, Eqs.~(3) and (4) become
\be
\left(\matrix{|\psi^k(8)_S\rangle\cr
|\psi^k(8)_A\rangle\cr
|\psi^k(10)_{S_t}\rangle\cr}\right)\quad = \quad \left(
\matrix{{1\over\sqrt 3} & 0 & -{2\over\sqrt 6}\cr
0 & 1 & 0\cr
{\sqrt{2}\over\sqrt 3} & 0 &{1\over\sqrt 3}\cr}\right)\quad\left(
\matrix{|\{q_a~q_b\}~q_a\rangle\cr
|[q_a~q_b]~q_a\rangle\cr
|q_a~q_a~q_b\rangle\cr}\right)~,
\en
and
\be
\left( \matrix{|\psi^\Sigma(8)_S\rangle\cr
|\psi^\Sigma(8)_A\rangle\cr
|\psi^\Lambda(8)_S\rangle\cr
|\psi^\Lambda(8)_A\rangle\cr
|\psi^{\Lambda_1}(1)_A\rangle\cr
|\psi^\Sigma(10)_{S_t}\rangle\cr}\right) \quad =\quad
\left( \matrix{
{1\over\sqrt{6}}& 0 & {1\over\sqrt{
6}}& 0 & {-2\over\sqrt{6}}& 0 \cr
0 & {-1\over\sqrt 2}& 0 & {-1\over\sqrt 2}&0&0\cr 
{1\over\sqrt 2}& 0 & {-1\over\sqrt 2}& 0 &0&0\cr 
0 & {1\over\sqrt{6}}& 0
&{-1\over\sqrt{6}}& 0 &{-2\over\sqrt{6}}\cr
0 &{1\over\sqrt 3}& 0 &{1\over\sqrt
3}& 0 &{1\over\sqrt 3}\cr
{1\over\sqrt 3}& 0 &{1\over\sqrt 3}& 0
&{1\over\sqrt 3}& 0 \cr}\right)\quad \left(
\matrix{|\{sd\}u\rangle&\cr |[sd]u\rangle&\cr |\{su\}d\rangle&\cr
|[su]d\rangle&\cr |\{du\}s\rangle&\cr
|[du]s\rangle&\cr}\right).
\en
Likewise, in this basis the identity matrix
becomes
\be
\hat{I}=\sum_{q_a,q_b,q_c}\Big(|\{q_a~q_b\}q_c
\ra\la\{q_a~q_b\}q_c|+|[q_a~q_b]q_c\ra\la[q_a~q_b]q_c|\Big)~.
\en
Then Eqs.(8)-(11) can be recast into the following form:
\be
|\psi^k(8)_A\rangle &=& \sum_{q_a,q_b,q_c}|[q_a~q_b]\,q_c\rangle~\langle 
[q_a~q_b]~q_c~|\psi^k(8)_A\rangle~,   \\
|\psi^k(8)_S\rangle &=& \sum_{q_a,q_b,q_c}|\{q_a~q_b\}~q_c\rangle~\langle 
\{q_a~q_b\}~q_c~|\psi^k(8)_S\rangle~,   \\
|\psi^k(1)_A\rangle &=& \sum_{q_a,q_b,q_c}|[q_a~q_b]~q_c\rangle~\langle 
[q_a~q_b]~q_c~|\psi^k(1)_A\rangle~,    \\
|\psi^k(10)_{S_t}\rangle &=&
\sum_{q_a,q_b,q_c}|\{q_a~q_b\}~q_c\rangle~\langle 
\{q_a~q_b\}~q_c~|\psi^k(10)_{S_t}\rangle~,   
\en
where we have used $\la[q_a~q_b]q_c|\psi^k(8)_S\ra=0$ and $\la\{q_a~q_b\}q_c|
\psi^k(8)_A\ra=0$. The coefficients on the right-hand side (r.h.s.) of Eqs.
(19)-(22) are 
the matrix elements in Eqs.~(16) and (17).

Here we would like to emphasize that it is  conceptually and practically
simpler to consistently 
use the orthonormal quark states as the
basis so that the identity operator has the simple expressions of Eq.~(12) or
Eq.~(18).  They
provide the proper transformation from the particle states to the quark
states and vice versa as
given by Eqs.~(3) and (4), equivalently by Eq.~(8) to (11); or Eqs.~(16) and
(17), equivalently
Eqs.~(19) to (22).  These are the crucial relations we shall use in
converting decay amplitudes in
terms of  particles to decay amplitudes in terms of quarks, i.e., the quark
diagram amplitudes.

Next we can form irreducible representations for the spin part of 
the particle from the tri-${1\over 2}$-spin states
\be
\pmatrix{|\chi^{\pm{1\over 2}}({1\over 2})_S\rangle\cr
|\chi^{\pm{1\over 2}}({1\over 2})_A\rangle\cr
|\chi^{\pm{1\over 2}}({3\over 2})_{S_t}\rangle\cr}=
\pmatrix{{1\over\sqrt 6}&{1\over\sqrt 6}&-{2\over\sqrt 6}\cr
{1\over\sqrt 2}&-{1\over\sqrt 2}&0\cr
{1\over\sqrt 3}&{1\over\sqrt 3}&{1\over\sqrt 3}\cr}
\pmatrix{|\pm{ 1\over 2}&\mp{ 1\over 2}&\pm{ 1\over 2}\rangle\cr
|\mp{ 1\over 2}&\pm{ 1\over 2}&\pm{ 1\over 2}\rangle\cr
|\pm{ 1\over 2}&\pm{ 1\over 2}&\mp{ 1\over 2}\rangle~ \cr},
\en
giving 6 equations; and
\be
|\chi^{\pm{3\over 2}}({3\over 2})_{S_t}\rangle = 
|\pm{ 1\over 2}~\pm{ 1\over 2}~\pm{ 1\over 2}\rangle~,
\en
giving 2 diagonal ones; totalling 8 equations.  The inverse of these
equations is also easy to write out.

The baryon states must be totally antisymmetric in interchanging the 
composing quarks.  Since the color part (which we do not discuss here, see 
e.g., Ref.~[12]) is antisymmetric, the product of the flavor and 
the spin parts must be symmetric as the spatial wave function is symmetric
for low-lying baryons. The decuplet baryons are made out of
\be
|B^{m,k}(10)\rangle~=~~|\chi^m({3\over 
2})_{S_t}\rangle~|\psi^k(10)_{S_t}\rangle~,~~~m=\pm{1\over 2},
\pm{3\over 2},~ \hbox{and} ~k=1~{\rm to}~10~.
\en
The octet baryon is a combination of two parts
\be
|B^{m,k}(8)\rangle &=& a~|B^{m,k}_A(8)\rangle~+~b~|B^{m,k}_S(8)\rangle \non\\
&=& a~|\chi^m({1\over 2})_A\rangle~|\psi^{k}(8)_A\rangle~+~b~|\chi^m({1\over 
2})_S\rangle~|\psi^k(8)_S\rangle~,
\en
where
\be
|a|^2~+~|b|^2~=~1~.
\en
The precise values of $a$ and $b$ are not known. 
Importantly, our formalism does
not need such information. We clearly show that the quark-diagram scheme is
independent of the values of $a$ and $b$.
 
If one assumes the SU(6) symmetry, then
$a=b={1\over\sqrt 2}$ and the octet baryon wave function can be rewritten as
\be
|B^{m,k}(8)\rangle={\sqrt{2}\over 3}\left(|\chi^m_{A_{12}}(\half)\ra
|\psi^k_{A_{12}(8)}\ra
                                         +|\chi^m_{A_{13}}(\half)\ra
|\psi^k_{A_{13}(8)}\ra
                                         +|\chi^m_{A_{23}}(\half)\ra
|\psi^k_{A_{23}(8)}\ra
\right),
\en
where the subscripts $ij$ to $A$ and $S$  indicate the pair that is 
antisymmetric or symmetric.  However, 
SU(6) is not an exact symmetry. It is important to show that the
quark-diagram scheme does not
depend on it.  Our formulation indeed confirms this requirement.

 Earlier Kohara had given a quark-diagram formulation for the
quark-mixing-allowed decays of the antitriplet charmed baryons [10].
For $B_c(\bar{3})\to B(8)+M(8)$ decays (of which he
considered only the
quark-mixing-allowed decays), it seems that  Kohara [10] was motivated by
Eq.~(28), which is true only if SU(6) symmetry is exact, and  chose the
octet baryon state to be an equal combination of two different pairs  of
quarks both antisymmetric
in  flavor and spin, i.e. two of the terms in Eq.~(28), which one
can easily check that the two terms are not orthonormal and they are not the
conventional way of making the octet baryon wave function. Choosing 
non-orthonormal basis for
the quark states, if done correctly, is all right though complicated.
[Kohara's
paper did not provide enough detailed
information in his paper for us to check directly his
calculations without redoing
all his formulation in the non-orthonormal basis he used.  However, we can
compare our results with
his for the part he had calculated.  Indeed our results agree  (see detailed
comparisons given
later by Eq.~(63).]   However,  we would still  like to emphasize that the
use of orthonormal basis
is much more convenient and, very importantly, that we have made sure that
the
quark-diagram scheme
does not depend upon SU(6) symmetry at all. For the $B_c(\bar{3})\to
B(10)+M(8)$
decays (of
which Kohara also only considered the mixing-matrix-allowed decays), Kohara
used the same basis as we use and our results agree. 

  Besides the $|B^{m,k} (8)\rangle$ states as given by Eq.~(26), there are 
the states orthogonal to them, which are denoted by
\be|B^{m,k}_\bot (8)\rangle~=~b^*~|\chi^m({1\over 
2})_A\rangle~|\psi^k(8)_A\rangle~-~a^*~|\chi^m({1\over 
2})_S\rangle~|\psi^k(8)_S\rangle~, 
\en
and
$\langle B_\bot^{m,k}(8)~|B^{m,k}(8)\rangle~=~0~.$
Nature does not realize these states, but they are there in the formalism and
hence must be considered when completeness of these states is used.  

Likewise, we can formulate the meson case, which is much simpler than 
the baryon case.  We discuss it here for completeness and for comparison.  
Mesons are made out of ${1\over 2}$-spin quark-antiquark $q'\bar q$ pair 
belonging to the flavor $[3] \times [\bar 3]$ representation.  They form
flavor irreducible 
representations of the $3\times \bar 3$ = 9 = 8 + 1, ~i.e., the 9 
quark-antiquark states can be decomposed into flavor [8] and [1] irreducible 
states denoted by $|\phi^j(8)\rangle$ and $|\phi(1)\rangle$ respectively,
where the superscript ``$j$'' denotes the eight particles in the [8] 
irreducible representations.

The transformation between the two bases, the quark basis and the 
irreducible-representation particle basis (both are orthonormal), 
can be written in a $9\times 9$ orthogonal matrix
\be
\pmatrix{|\phi^{\pi^+}\rangle&\cr
|\phi^{K^+}\rangle&\cr
|\phi^{\pi^-}\rangle&\cr
|\phi^{K^0}\rangle&\cr
|\phi^{K^-}\rangle&\cr
|\phi^{\bar{K}^0}\rangle&\cr
|\phi^{\pi^0}\rangle&\cr
|\phi^{\eta_8}\rangle&\cr
|\phi^{\eta_1}\rangle&\cr}
=
\pmatrix{1&0&0&0&0&0&0&0&0\cr
0&1&0&0&0&0&0&0&0\cr
0&0&1&0&0&0&0&0&0\cr
0&0&0&1&0&0&0&0&0\cr
0&0&0&0&1&0&0&0&0\cr
0&0&0&0&0&1&0&0&0\cr
0&0&0&0&0&0&{1\over\sqrt 2}&{-1\over\sqrt 2}&0\cr
0&0&0&0&0&0&{1\over\sqrt 6}&{1\over\sqrt 6}&{-2\over\sqrt 6}\cr
0&0&0&0&0&0&{1\over\sqrt 3}&{1\over\sqrt 3}&{1\over\sqrt 3}\cr}
\pmatrix{|u\bar d\rangle&\cr
|u\bar s\rangle \cr
|d\bar u\rangle \cr
|d\bar s\rangle \cr
|s\bar u\rangle \cr
|s\bar d\rangle \cr
|u\bar u\rangle \cr
|d\bar d\rangle \cr
|s\bar s\rangle \cr}.
\en
The nine equations given by the matrix equation can also be 
written out as
\be
|M^j(8)\rangle~=~\sum_{\bar q,q'}|\bar qq'\rangle~\langle \bar 
qq'~|~M^j(8)\rangle~,
\en
and
\be
|M(1)\rangle~=~\sum_{\bar q,q'}|\bar qq'\rangle~\langle \bar 
qq'~|~M(1)\rangle~,
\en
where the summation is for $\bar q=\bar u,\bar d,\bar s$ and $q'=u,d,s$. 
These equations are obtained simply by multiplying the left-hand side of 
(31) and (32) by 
\be
\hat I=\sum_{\bar q,q'}|\bar qq'\rangle~\langle \bar qq'|~,
\en
which is the completeness relation of the orthonormal
$|q'\bar q\rangle$-basis in the quark-antiquark vector space.

The irreducible-representation states in spin are related to the 
spin-product space by
\be
\left(\matrix{
|\chi^+(1)\rangle\cr
|\chi^-(1)\rangle\cr
|\chi^0(1)\rangle\cr
|\chi(0)~\rangle\cr}\right)=
\left(\matrix{
1&0&0&0\cr
0&1&0&0\cr
0&0&{1\over\sqrt 2}&{1\over\sqrt 2}\cr
0&0&{1\over\sqrt 2}&{-1\over\sqrt 2}\cr}
\right)
\left(\matrix{
|{+1\over 2}~{+1\over 2}\rangle\cr
|{-1\over 2}~{-1\over 2}\rangle\cr
|{-1\over 2}~{+1\over 2}\rangle\cr
|{+1\over 2}~{-1\over 2}\rangle\cr}
\right)
\en
and its inverse is trivially obtained (by using the transpose of the matrix)
\be
\left(\matrix{
|{+1\over 2}~{+1\over 2}\rangle\cr
|{-1\over 2}~{-1\over 2}\rangle\cr
|{-1\over 2}~{+1\over 2}\rangle\cr
|{+1\over 2}~{-1\over 2}\rangle\cr}
\right)
=
\left(\matrix{
1&0&0&0\cr
0&1&0&0\cr
0&0&{1\over\sqrt 2}&{1\over\sqrt 2}\cr
0&0&{1\over\sqrt 2}&{-1\over\sqrt 2}\cr}\right)
\left(\matrix{
|\chi^+(1)\rangle\cr
|\chi^-(1)\rangle\cr
|\chi^0(1)\rangle\cr
|\chi(0)~\rangle\cr}\right).
\en
For pseudoscalar mesons, the wave functions are simply given by
\be
|M(1)~\rangle &=& |\chi(0)\rangle~|\phi(8)~\rangle~, \non \\
|M^j(8)\rangle &=& |\chi(0)\rangle~|\phi^j(8)\rangle~,
\en
where the superscript $j$ indicates the eight different particles
given in Eq.~(30).
\vskip 0.3cm
\pagebreak
\noindent {\bf III. Quark Diagram Scheme for $B_c(\bar 3) \to B(10) +M(8) $}

  The light quarks of the charmed baryons belong to either a $[\bar{ 3}]$ or
a $[6]$ representation of the flavor SU(3). The $\Lambda_c^+$,
$\Xi_c^{+A}$, and $\Xi_c^{0A}$ form a
 $[\bar { 3}]$ representation. They all decay weakly.  The $\Omega_c^0$,
$\Xi_c^{+S}$, $\Xi_c^{0S}$,
 $\Sigma_c^{++}$, $\Sigma_c^{+}$, $\Sigma_c^{0}$ form a [6]
representation; among them,
however, only  $\Omega_c^0$  decays weakly (the $\Sigma_c^{++,+,0}$ decay
strongly to the $\Lambda_c^+$ of the 
[$\bar{ 3}$] representation and the $\Xi_c^{+,0}$ decay electromagnetically).

We shall first discuss the simpler case of the 
decuplet baryon being in the decay products.
\pagebreak

\noindent {\bf III.a. Formalism}

Consider a particular charmed baryon $B^{m,{i_0}}_c$ decaying into an octet
meson 
$M^{j_0}(8)$ and a decuplet baryon $B^{m,{k_0}}(10)$, where the 
subscript ``0'' signifies that we are discussing a specific baryon and a 
specific meson.  The amplitude with the spin-projection ${m,m'}$ 
summed over is 
\be
A(i_0 \rightarrow j_0~k_0) 
&\equiv& \sum_{m,m'} \langle B^{m,i_0}_c~|\hat 
H_W|M^{j_0}\rangle~|B^{m',{k_0}}(10)\rangle{\rm ~;~ using~ Eq.~(25)~ for~} 
|B^{m,{k_0}}(10)\rangle    \non \\
&=& \sum_{m,m'}\langle B^{m,{i_0}}_c~|\hat
H_W|M^{j_0}(8)\rangle~|\chi^{m'}({3\over 
2})_{S_t}\rangle~|\psi^{k_0}(10)_{S_t}\rangle{\rm ;~inserting~Eq.(12),}
\non\\
&=& \sum_{m,m',{q_i}}\langle B^{m,{i_0}}_c~|\hat 
H_W|M^{j_0}(8)\rangle~|\chi^{m'}({3\over 
2})_{S_t}\rangle~|q_1~q_2~q_3\rangle ~\langle 
q_1~q_2~q_3|\psi^{k_0}(10)_{S_t}\rangle~;   \non \\
&&~~~~~{\rm using~Eq.~(36)~for~}M^{j_0}(8)~,  \non  \\
&=& \sum_{m,m',{q_i}}\langle B^{m,{i_0}}_c~|\hat 
H_W|\chi(0)\rangle~|\phi^{j_0}(8)\rangle ~|\chi({3\over 2})_{S_t}\rangle~
|q_1~q_2~q_3\rangle~\langle 
q_1~q_2~q_3|~\psi^{k_0}(10)_{S_t}\rangle~;   \non \\
&&~~~~~{\rm inserting~Eq.~(33),}    \non \\
&=& \sum_{m,m',\bar q,q',q_i}\langle  
B^{m,i_0}_c|\hat H_W|\chi(0)\rangle ~|\chi^{m'}({3\over
2})_{S_t}\rangle~|\bar q
q'\rangle~|q_1~q_2~q_3\rangle~\langle \bar qq'|\phi^{j_0}(8)\rangle  \non \\
&&~~~~\times\langle  q_1~q_2~q_3~|\psi^{k_0}(10)_{S_t}\rangle \non \\
&\equiv& \sum_{\bar q,q',{q_i}} A(i_0\rightarrow \bar
q\,q'~q_1~q_2~q_3)~\langle
\bar q\,q'|\phi^{j_0}(8)\rangle~\langle 
q_1~q_2~q_3~|\psi^{k_0}(10)_{S_t}\rangle~, 
\en
where 
\be
A(i_0 \rightarrow \bar q\,q'~q_1~q_2~q_3)
\equiv\sum_{m,m'}\langle
B^{m,{i_0}}_c|\hat 
H_W|\chi(0)\rangle~|\chi^{m'}({3\over 2})_{S_t}\rangle~|\bar
q\,q'~q_1~q_2~q_3\rangle
\en
are the quark-diagram amplitudes.  
Therefore, Eq.~(37) gives the particle amplitudes $A(i_0 \rightarrow
j_0~k_0)$ of 
$B^{i_0}_c$ decaying into particles $M^{j_0}(8)$ and $B^{k_0}(10)$ 
in terms of the quark amplitudes $A(i_0 \rightarrow \bar q\,q'~q_1~q_2~q_3)$
of $B^{i_0}_c$ decaying
into quarks 
$\bar q\,q'~q_1~q_2~q_3$.  The coefficients $\langle \bar q
q'|\phi^{j_0}(8)\rangle$ and $\langle 
q_1~q_2~q_3|\psi^{k_0}(10)_{S_t}\rangle$, each set of which forming elements
of an orthogonal
matrix,  are those given in Eq.~(30) and  Eqs.~(3) to (7), respectively. 

Using the orthonormality of the coefficients,   
we can easily convert Eq.~(37) to express the quark amplitudes in terms 
of the particle amplitudes
\be
A(i_0 \rightarrow \bar q\,q'~q_1~q_2~q_3)=\sum_{j_0,k_0} A(i_0 \rightarrow 
{j_0}~{k_0})~\langle \phi^{j_0}(8)|\bar q q'
\rangle~\langle q_1q_2q_3|\psi^{k_0}(10)_{S_t}\rangle~,
\en
using the orthonormality condition of the coefficients, which 
is the result of the orthonormality of the states.

We can also formulate the relation (37) in the basis given by Eqs.~(13)
and (14), which is also more convenient to apply since
$|\psi^{k_0}(10)_{S_t}\ra$ is totally symmetric.
Replacing ``inserting Eq.~(12)" by ``inserting Eq.~(18)" in Eq.~(37), we
obtain 
\be
A(i_0\ri j_0~k_0)
\equiv\, \sum_{\bar{q},q',q_i}A(i_0\ri\bar{q}\,q'~\{q_a~q_b\}q_c)\la
q'\bar{q}|\phi^{j_0}(8)\ra\la\{q_a~q_b\}q_c|\psi^{k_0}(10)_{S_t}\ra~,  
\en
where
\be
A(i_0\rightarrow\bar q q',\{q_a~q_b\}q_c)\equiv\sum_{m,m'}\langle  
B^{m,i_0}_c|\hat H_W|\chi(0)\rangle ~|\chi^{m'}({3\over
2})_{S_t}\rangle~|\bar 
qq',\{q_a~q_b\}q_c\rangle~.
\en
Let us look more carefully at the amplitudes.
For $q_a=q_b$,
\be
A(i_0\rightarrow\bar q q',\{q_a~q_b\}q_c) &=& A(i_0\rightarrow \bar q 
q',q_a~q_a~q_c)   \non \\ 
&\equiv& A_S\Bigl( B_c(\bar 3)\rightarrow B(10)~M(8)\Bigr),  
\en
and for $q_a\ne q_b$,
\be
A(i_0\rightarrow\bar q q',\{q_a~q_b\}q_c) &=&
{1\over \sqrt 2}[A(i_0\rightarrow\bar q q',q_a~q_b~q_c)
+A(i_0\rightarrow\bar q q',q_b~q_a~q_c)]  \non \\
&\equiv& \sqrt 2~A_S\Bigl(B_c(\bar 3)\rightarrow 
B(10)~M(8)\Bigr), 
\en
where we have used 
\be
A(i_0\rightarrow\bar q q',q_a~q_b~q_c)
&=& {1\over 2}[A(i_0\rightarrow\bar q q',q_a~q_b~q_c)+A(i_0\rightarrow\bar 
q q',q_b~q_a~q_c)]_{q_a\ne q_b}  \non \\
&\equiv& A_S\Bigl(B_c(\bar 3)\rightarrow B(10)~M(8)\Bigr)~.
\en
We shall see later that this assumption gives results consistent 
with those using the SU(3)-IR amplitudes.
Eqs.~(42) and (43) can be combined into one equation
\be
A(i_0\rightarrow\bar q q',\{q_a~q_b\}q_c)=[\sqrt 2 (1-\delta_{q_aq_b})+
\delta_{q_aq_b}]~A_S\Bigl(B_c(\bar 3)\rightarrow 
B(10)~M(8)\Bigr)~, \non
\en
which we substitute into Eq.~(40) and obtain

\be
A(i_0\ri j_0~k_0)
&\equiv& \sum_{\bar{q},q',q_i} [\sqrt 2 (1-\delta_{q_aq_b})+
\delta_{q_aq_b}]~A_S\Bigl(B_c(\bar 3)\rightarrow B(10)~M(8)\Bigr)   \non \\
&\times&
\la q'\bar{q}|\phi^{j_0}(8)\ra\la\{q_a~q_b\}q_c|\psi^{k_0}(10)_{S_t}\ra~. 
\en
Here in Eq.~(37) and in Eq.~(45) we see the important use of  Eq.~(12) and of
Eq.~(18) to convert particle-amplitudes to the quark-amplitudes.

   One can easily show by diagram drawing that the $B_c(\bar{3})\ri
B(10)+M(8)$ decays have contributions  only from the $W$-exchange and the 
horizontal $W$-loop diagrams, i.e., the $\c$ and $\e$
types of amplitudes. In the
$\a$ and $\b$ amplitudes, the two spectator quarks that are antisymmetrized
in the initial charmed
baryon state remain to be  antisymmetrized after the weak-interaction
decay and cannot contribute to make an
$B(10)$ whose wave function is totally symmetric. In the $\c$ and $\e$ types
of amplitudes, an
appropriate  quark pair $\bar{q}_0q_0$ is created so that the $\bar{q}_0$
will combine with one of the quarks originated from the initial quark to form
the meson $j_0$.
Depending upon where the   pair $\bar{q}_0q_0$ can be inserted in the
diagrams, we have different
types of  $\c_S$ and $\e_S$ of amplitudes; $\c_{1S}$ for $\bar{q}_0$ forming
a meson with a spectator quark (which does not contribute in this case of 
$B(10)$ in the final state);  $\c_{2S}$
for $\bar{q}_0$ forming a meson with the weak-interacting non-charmed quark;
$\c'_S$ for
$\bar{q}_0$ forming a meson with the quark decayed from the charmed
quark; $\e_S$ for $\bar{q}_0$ forming a meson with the a spectator quark; 
and $\e'_S$ for $\bar{q}_0$ forming a
meson with the quark decayed from the charmed quark. The quark
$q_0$ from the pair creation will form with the other two quarks to become
the final baryon $k_0$. Thus in Eq.~(45) only $q_1$ and $q_2$ are summed 
over and Eq.~(45) becomes
\be
 A(i_0 &\ri& j_0k_0)  \\
&=& \cts\Bigl(B_c(\bar 3)\rightarrow B(10)M(8)\Bigr)
[\sqrt 2(1-\delta_{q_1q_3})+\delta_{q_1q_3}]\la\bar{q}_0q'_2|\phi^{j_0}(8)
\ra\la\{q_1~q_3\}q_0|\psi^{k_0}(10)_{S_t}\ra  \non \\
&+& \csp\Bigl(B_c(\bar 3)\rightarrow B(10)M(8)\Bigr)
[\sqrt 2(1-\delta_{q_1q'_2})+\delta_{q_1q'_2}]\la\bar{q}_0q_3|\phi^{j_0}(8)
\ra\la\{q_1~q'_2\}q_0|\psi^{k_0}(10)_{S_t}\ra \non \\
&+& \es\Bigl(B_c(\bar 3)\rightarrow B(10)M(8)\Bigr)
[\sqrt 2(1-\delta_{q_3q_1})+\delta_{q_3q_1}]
\la\bar{q}_0q_2|\phi^{j_0}(8)\ra\la\{q_3~q_1\}q_0|\psi^{k_0}(10)_{S_t}\ra.\non
\en

 Using (46) for $B_c(\bar 3)\ri B(10)+M(8)$ decays, we obtain column two of
Tables
1a, 1b and 1c. (In these tables we have dropped the parenthesis that 
specify the decay of $B_c(\bar 3)\rightarrow B(10)~M(8)$.)  We see that 
all $B_c(\bar 3)\rightarrow B(10)~M(8)$ decays, fifty-five of them, can 
be expressed in terms of the three unknown amplitudes: $\cts,~\csp$ and 
$\e_S$. Therefore, we obtain many relations among the particle decay
amplitudes as shown in the next section.

Next we make an attempt to include  and parametrize the effects of
final-state interactions and
SU(3) symmetry breaking. For final-state interactions we introduce an
explicit factor
$e^{i\delta}$ to each isospin partial-wave amplitude. These phase shifts
$\delta$'s in general have both real and imaginary parts; the imaginary
component indicates the inelastic effect.  SU(3)-violation
effects can manifest in several places. For example, the quark diagrams 
with $s\bar s$ insertion are {\it a priori} different from those arising 
from $u\bar u$ or $d\bar d$ insertion. However, for the simplicity of the
presentation of the  Tables in the present  paper, we parametrize SU(3)
symmetry breaking only in  the
$\e$  type of quark diagrams whose presence  (i.e. contributions to the
quark-mixing-singly-suppressed  decay modes) are solely due to SU(3)
violation
effects.

\vskip 0.3in
\noindent {\bf III.b. Results and Tables}

  In the absence of effects from SU(3) breaking and final-state interactions,
the following relations can be obtained from the second column of Tables 1a
to 1c, namely 
\be
 |A(\Lambda_c^+ \to \Sigma^{*+} \eta_8)|^2 &=& |A(\Xi^{0A} \to \Xi^{*0} 
\eta_8)|^2,   \non 
\en
\be
 |A(\Xi_c^{0A} \to \Omega^- K^+)|^2 &=& 3|A(\Xi_c^{0A} \to \Xi^{*-} 
\pi^+)|^2=3|\Lambda_c^+\to\Xi^{*0}K^+)|^2  \non \\
 =6|A(\Xi_c^{0A}\to\Xi^{*0}\pi^0)|^2 &=& 6|A(\Lambda_c^+\to\Sigma^
{*+}\pi^0)|^2=6|A(\Lambda_c^+\ri\Sigma^{*0}\pi^+)|^2,  
\en
\be
 |A(\Lambda_c^+\to\Delta^{++}K^-)|^2 &=& 3|A(\Lambda_c^+\to\Delta^+
\bar{K}^0)|^2  \non \\
 =3|A(\Xi_c^{0A}\to\Sigma^{*+}K^-)|^2 &=& 6|A(\Xi_c^{0A}\to\Sigma^
{*0}\bar{K}^0)|^2  \non
\en
for quark-mixing-allowed modes; 
\be
 |A(\Xi_c^{0A}\to\Sigma^{*0}\pi^0)|^2 =
3|A(\Xi_c^{0A}\to\Sigma^{*0}\eta_8)|^2,   \non 
\en
\be
 |A(\Xi_c^{0A}\to\Sigma^{*-}\pi^+)|^2 &=& |A(\Xi_c^{0A}\to\Xi^{*-}K^+)
|^2=4|A(\Lambda_c^+\to\Delta^0\pi^+)|^2  \non \\
 =4|A(\Xi_c^{+A}\to\Xi^{*0}K^+)|^2 &=& 8|A(\Lambda_c^+\to\Sigma^{*0}K^+)|^2 
\non \\
 =8|A(\Xi_c^{+A}\to\Sigma^{*0}\pi^+)|^2 &=& 8|A(\Xi_c^{+A}\to\Sigma^{*+}
\pi^0)|^2,    
\en
\be
 |A(\Lambda_c^+\to\Delta^{++}\pi^-)|^2 &=& |A(\Xi_c^{+A}\to\Delta^{++}
K^-)|^2=3|A(\Lambda_c^{+}\to\Sigma^{*+}K^0)|^2  \non \\
 =3|A(\Xi_c^{+A}\to\Delta^+\bar{K}^0)|^2 &=& 3|A(\Xi_c^{0A}\to\Delta^0
\bar{K}^0)|^2=3|A(\Xi_c^{0A}\to\Xi^{*0}K^0)|^2  \non \\
 =3|A(\Xi_c^{0A}\ri\Sigma^{*+}\pi^-)|^2 &=& 3|A(\Xi_c^{0A}\to\Delta^+K^-)|^2
\non   
\en
for quark-mixing-suppressed modes; 
\be
|A(\Xi_c^{+A} \to \Delta^+ \eta_8)|^2 &=& |A(\Xi_c^{0A} \to \Delta^{0}
\eta_8)|^2,   \non \\
|A(\Xi_c^{0A}\to\Delta^0\pi^0)|^2 &=& 2|A(\Xi_c^{+A}\to\Delta^+\pi^0)|^2,
\non
\en
\be
 |A(\Xi_c^{+A}\to\Delta^{++}\pi^-)|^2 &=& 3|A(\Xi_c^{+A}\to\Sigma^{*+}
K^0)|^2  \non \\
 =3|A(\Xi_c^{0A}\to\Delta^+\pi^-)|^2 &=& 6|A(\Xi_c^{0A}\ri\Sigma^{*0}K^0)|^2,
\en
\be
|A(\Xi_c^{0A}\to\Delta^-\pi^+)|^2 &=& 3|A(\Xi_c^{+A}\to\Delta^0\pi^+)
|^2 \non \\
 =3|A(\Xi_c^{0A}\to\Sigma^{*-}K^+)|^2 &=& 6|A(\Xi_c^{+A}\to\Sigma^{*0}
K^+)|^2 \non 
\en
for  quark-mixing-doubly-suppressed modes, and many
relations between quark-mixing-allowed, -suppressed, and -doubly-suppressed
decay modes, for example
\be
|A(\Lambda_c^+ \to \Delta^0 \pi^+)|^2 &=& 2s_1^2|A(\Lambda_c^+ \to
\Sigma^{*+}
\pi^0)|^2,   \non \\
|A(\Lambda_c^+ \to \Delta^{++} \pi^-)|^2 &=& s_1^2|A(\Lambda_c^+ \to
\Delta^{++}K^-)|^2,   \non \\
|A(\Xi_c^{+A} \to \Sigma^{*0} K^+)|^2 &=& s_1^4|A(\Lambda_c^+ \to
\Sigma^{*+} \pi^0)|^2,   \\
|A(\Xi_c^{+A} \to \Delta^{++} \pi^-)|^2 &=& s_1^4|A(\Lambda_c^+ \to
\Delta^{++} K^-)|^2. \non
\en

  Several comments are in order. (i) Table~1.a and relations (47) for
quark-mixing-allowed decays $B_c(\bar 3)\to B(10)+M(8)$ are also obtained
previously by Kohara [10] and are in agree with ours. (ii)
Though all the above quark-diagram relations are obtained
in the absence of effects from SU(3) violations and final-state interactions,
however, if each quark-diagram amplitude in the SU(3) relation contains only
a single isospin phase shift, then such a relation holds even in the presence
of final-state interactions because of SU(3) symmetry for phase shifts.
Examples are the first relation in Eqs.~(47) and (49). The same is true for
the relations obtained below. (iii) We
note that the quark-mixing-allowed decays of an antitriplet
charmed baryon into a decuplet baryon and a pseudoscalar meson can occur only
through ${\it W}$-exchange diagrams.
The experimental measurement of $\Lambda_c^+ \to \Delta^{++} K^- $
[2] indicates that the ${\it W}$-exchange mechanism
plays a significant role in charmed baryon decays. 
(iv) The quark-mixing-allowed decays of $\Xi_c^{+A}$ 
and quark-mixing-doubly-suppressed decays of $\Lambda_c^+$ into a
decuplet baryon are prohibited in the quark-diagram scheme:
\be
|A(\Xi_c^{+A}\to \Sigma^{*+}\bar{K}^0)|^2 &=& 0,~~~|A(\Xi_c^{+A}\to \Xi^{*0}
\pi^+)|^2=0,  \non \\
|A(\Lambda_c^+\to \Delta^+K^0)|^2 &=&
0,~~~|A(\Lambda_c^+\to\Delta^0K^+)|^2=0.
\en
In the SU(3)-IR  approach of Savage and Springer (SS) [6], these 
decays are
governed by the reduced matrix element $\alpha$ defined in Eq.~(17) of 
Ref.~[6]. However, we see that they are forbidden in the quark-diagram scheme
since they are given by the quark diagram ${\cal A}$ or ${\cal B}'$ and 
they give zero contribution, as
we discussed before, because of the un-matching symmetry properties of the
antitriplet charmed
baryon  and the decuplet baryon. Furthermore, we note that the
SU(3)-IR approach of SS will predict the above relations
(48-51) only if the reduced matrix elements $\alpha$ and $\gamma$ make
no contributions.
As a consequence, there are only two independent SU(3) reduced matrix
elements $\beta$ and $\delta$. The quark-diagram amplitudes
and the SU(3)-symmetry parameters are related by
\be
\beta=\, {1\over 2}({\cal C}'_S + {\cal C}_{2S} ),~~\delta=\, {1\over 2}
({\cal C}'_S - {\cal C}_{2S} ),~~\alpha= \gamma=0.
\en

\vskip 0.3cm
\pagebreak
\noindent {\bf IV.  Quark Diagram Scheme for $B_c(\bar 3)\rightarrow
B(8)+M(8)$}

\noindent {\bf IV.a.  The Formalism}

The formalism is very similar to that given in Sect. III.a. for the
decuplet baryon in the final state except for
the complication that the octet baryons are made up with two orthonormal 
parts, Eq.~(26).  We shall see that all it does is that each type of the
quark
amplitude $A$ will be made up of two independent ones, the symmetric and the
antisymmetric.
Following the similar procedure used in Eqs.~(37) and (50), we derive
\be
A(i_0\rightarrow j_0~k_0) &=& \sum_{m,m'}\langle B^{m,i_0}_c|\hat 
H_W|M^{j_0}(8)\rangle~|B^{m',k_0}(8)\rangle   \non 
\en
\be
~~~~~~&=& \sum_{m,m'}\langle B^{m,i_0}_c|\hat 
H_W|M^{j_0}(8)\rangle~\Big(a~|\chi^{m'}({1\over 2})_A\rangle 
|\psi^{k_0}(8)_A\rangle + b~|\chi^{m'} 
({1\over 2})_S\rangle |\psi^{k_0}(8)_S\rangle\Big) \non   \\
 &=& \sum_{m,m',q_i} a\,\langle B^{m,i_0}_c| 
H_W|M^{j_0}(8)\rangle~|\chi^{m'}({1\over 2})_A\rangle~|q_1~q_2~q_3\rangle~
\langle q_1~q_2~q_3|\psi^{k_0}(8)_A\rangle   \non \\
&+& \sum_{m,m',q_i} b\,\langle B^{m,i_0}_c| 
H_W|M^{j_0}(8)\rangle~|\chi^{m'}({1\over 2})_S\rangle~|q_1~q_2~q_3\rangle~
\langle q_1~q_2~q_3|\psi^{k_0}(8)_S\rangle  \non \\
 &=& \sum_{m,m',q_i} a\,\langle B^{m,i_0}_c| 
H_W|M^{j_0}(8)\rangle~|\chi^{m'}({1\over 2})_A\rangle~|[q_1~q_2]~q_3\rangle~
\langle [q_1~q_2]~q_3|\psi^{k_0}(8)_A\rangle   \non \\
&+& \sum_{m,m',q_i} b\,\langle B^{m,i_0}_c| 
H_W|M^{j_0}(8)\rangle~|\chi^{m'}({1\over
2})_S\rangle~|\{q_1~q_2\}~q_3\rangle~
\langle \{q_1~q_2\}~q_3|\psi^{k_0}(8)_S\rangle.~~  
\en

To decompose the meson state  
into the  $q'\bar q$ state, we insert in Eq.~(53) the  completeness 
relation Eq.~(33) and obtain
\be
A(i_0\rightarrow j_0~k_0) &=& \!\!\!\!\!\!\!\sum_{m,m',\bar q,q',q_i}~\!\! 
b^*\,\langle B^{m,i_0}_c 
|\hat H_W|\chi(0^-)\rangle~|\bar q q'\rangle~|\chi^{m'}({1\over 
2})_A\rangle |[q_1~q_2]~q_3\rangle    \non \\
&&~~~~~~~~~~\times\langle\bar 
qq'|\phi^{j_0}(8)\rangle~\langle [q_1~q_2]~q_3|\psi^{k_0}(8)_A\rangle   \non
\\
&-& \!\!\!\!\!\!\sum_{m,m',\bar q,q',q_i} a^*\,\langle B^{m,i_0}_c 
|\hat H_W|\chi(0^-)\rangle~|\bar q q'\rangle~|\chi^{m'}({1\over 
2})_S\rangle~|\{q_1~q_2\}~q_3\rangle   \non  \\
&&~~~~~~~~~~\times\langle \bar q 
q'|\phi^{j_0}(8)\rangle~\langle {q_1~q_2}~q_3|\psi^{k_0}(8)_S\rangle \non  \\
&\equiv& \!\sum_{\bar q,q',q_i}~A(i_0\rightarrow \bar
q~q'~[q_1~q_2]~q_3)~\langle 
\bar qq'|\phi^{j_0}(8)\rangle~\langle 
[q_1~q_2]~q_3|\psi^{k_0}(8)_A\rangle   \non \\
&+& \!\sum_{\bar q,q',q_i}~A(i_0\rightarrow \bar
q~q'~\{q_1~q_2\}~q_3)~\langle 
\bar qq'|\phi^{j_0}(8)\rangle~\langle 
\{q_1~q_2\}~q_3|\psi^{k_0}(8)_S\rangle~,    
\en
where 
\be
A(i_0\rightarrow \bar q~q'~[q_1~q_2]~q_3) &\equiv&
\!\!\!\!\!\!\!\sum_{m,m'}~\!\! 
b^*\,\langle B^{m,i_0}_c  |\hat H_W|\chi(0^-)\rangle~|\bar q
q'\rangle~|\chi^{m'}({1\over 
2})_A\rangle |[q_1~q_2]~q_3\rangle    \non \\
&=& A_A(B_c(\bar 3)\rightarrow B(8) M(8)),
\en 
and
\be
A(i_0\rightarrow \bar q~q'~\{q_1~q_2\}~q_3) &\equiv&
\!\!\!\!\!\!\!\sum_{m,m'}~\!\! 
b^*\,\langle B^{m,i_0}_c  |\hat H_W|\chi(0^-)\rangle~|\bar q
q'\rangle~|\chi^{m'}({1\over 
2})_A\rangle |\{q_1~q_2\}~q_3\rangle    \non \\
&=& [\sqrt 2(1-\delta_{q_1q_2})+\delta_{q_1q_2}]A_S(B_c(\bar 3)\rightarrow
B(8) M(8)).
\en
Now the decay amplitudes into particles are
related to decay amplitudes into quarks.

Therefore, the important result we have established is that for the 
decays into the $B(8)$, the quark diagrams have two independent types: 
the symmetric and the antisymmetric, $A_A$ and $A_S$.  This result is
independent of what particles
the 
$B(8)$'s decay from or are associated with.

   Let us discuss now specifically what types of quark diagram amplitudes
will contribute. For
$B_c(\bar 3)\ri B(8)+M(8)$, the two initial non-charmed quarks, say $q_1$ and
$q_2$, are antisymmetric
in flavor. In diagram $\a$, $q_1$ and $q_2$ are spectators; therefore, they
stay
antisymmetric in the final state. We denote the quark arising from the
charmed 
quark decay as $q_3$, and the quark-antiquark pair from the $W$ as
$\bar{q}_0q'_0$.
In diagram $\b'$ (the superscript ``$'$'' signifies that the quark $q_3$
coming
from the charmed quark decay contributes to the final meson formation rather 
than the final baryon formation), $q_1$ and $q_2$ are also spectators; 
therefore, they stay antisymmetric in the final product. In diagram $\b$,
$q_3$ and $q'_0$ are forced to be flavor antisymmetric due to the Pati-Woo
theorem [11], so are the quark pair $q'_1q_3$ in diagram $\c_1$. 
Note that the 
quark-diagram amplitudes ${\cal B}'_S$ and ${\cal C}_{1S}$ vanish because
of the Pati-Woo theorem which results from the fact that the 
$(V-A)\times(V-A)$ structure of weak
interactions is invariant under the Fierz transformation and that the baryon
wave function is color antisymmetric.
This theorem requires that the quark pair in a baryon produced by weak 
interactions be antisymmetric in flavor. Putting together all these 
information and referring to Fig.~2, we find
the detailed expression for  Eq.~(54)
\be
A(i_0\ri j_0k_0) &=& \a_A\Bigl(B_c(\bar 3)\rightarrow 
B(8)M(8)\Bigr)\langle\bar q_0 q'_0|\phi^{j_0}(8)\rangle\langle 
[q_1q_2]q_3|\psi^{k_0}(8)_A\rangle~ \non   \\
 &+& \b'_A\Bigl(B_c(\bar 3)\rightarrow 
B(8)M(8)\Bigr)\langle\bar q_0 q'_3|\phi^{j_0}(8)\rangle\langle 
[q_1q_2]q'_0|\psi^{k_0}(8)_A\rangle~ \non    \\
 &+& \b_A\Bigl(B_c(\bar 3)\rightarrow 
B(8)M(8)\Bigr)\langle\bar q_0 q_2|\phi^{j_0}(8)\rangle\langle 
[q_1q_3]q'_0|\psi^{k_0}(8)_A\rangle~ \non   \\
 &+& \c_{1A}\Bigl(B_c(\bar 3)\rightarrow 
B(8)M(8)\Bigr)\langle\bar q_0 q_2|\phi^{j_0}(8)\rangle\langle 
[q'_1q_3]q_0|\psi^{k_0}(8)_A\rangle~ \non   \\
&+& \c_{2A}\Bigl(B_c(\bar 3)\rightarrow 
B(8)M(8)\Bigr)\langle\bar q_0 q'_2|\phi^{j_0}(8)\rangle\langle 
[q_1q_3]q_0|\psi^{k_0}(8)_A\rangle    \non \\
&+& \c_{2S}\Bigl(B_c(\bar 3)\rightarrow 
B(8)M(8)\Bigr) 
\langle\bar q_0 q'_2|\phi^{j_0}(8)\rangle\langle 
\{q_1q_3\}q_0|\psi^{k_0}(8)_S\rangle [\sqrt
2(1-\delta_{q_1q_3})+\delta_{q_1q_3}]   \non \\
 &+& \c'_A\Bigl(B_c(\bar 3)\rightarrow 
B(8)M(8)\Bigr)\langle\bar q_0 q_3|\phi^{j_0}(8)\rangle\langle 
\{q_1q'_2\}q_3|\psi^{k_0}(8)_A\rangle   \non \\
&+& \c'_S\Bigl(B_c(\bar 3)\rightarrow 
B(8)M(8)\Bigr)
\langle\bar q_0 q_3|\phi^{j_0}(8)\rangle\langle 
\{q_1q'_2\}q_0|\psi^{k_0}(8)_S\rangle [\sqrt
2(1-\delta_{q_1q'_2})+\delta_{q_1q'_2}]   \non  \\
 &+& \e_A\Bigl(B_c(\bar 3)\rightarrow 
B(8)M(8)\Bigr)\langle\bar q_0 q_2|\phi^{j_0}(8)\rangle\langle 
[q_3q_1]q_0|\psi^{k_0}(8)_A\rangle    \non \\
&+& \e_S\Bigl(B_c(\bar 3)\rightarrow 
B(8)M(8)\Bigr)
\langle\bar q_0 q_2|\phi^{j_0}(8)\rangle\langle 
\{q_3q_1\}q_0|\psi^{k_0}(8)_S\rangle [\sqrt
2(1-\delta_{q_3q_1})+\delta_{q_3q_1}]\non \\
 &+& \e'_A\Bigl(B_c(\bar 3)\rightarrow 
B(8)M(8)\Bigr)\langle\bar q_0 q_3|\phi^{j_0}(8)\rangle\langle 
[q_1q_2]q_0|\psi^{k_0}(8)_A\rangle.  
\en
Applying this to all the $B_c(\bar 
3)\rightarrow B(8)~M(8)$ decays, we can express all the 58 decays in terms of
the eleven unknown amplitudes in (57) (see also Table~2).

\vskip 0.15cm
\noindent{\bf IV.b. Results and Tables}

     From the second column of Tabs. 2a-2c, which gives the results in the
absence of effects from SU(3)
violations and final state interactions, we have the following relations :
\be
|A(\Xi_c^{0A} \to \Sigma^- \pi^+)|^2 &=& |A(\Xi_c^{0A} \to \Xi^- K^+)|^2,
\non  \\
|A(\Xi_c^{0A}\to n\bar{K}^0)|^2 &=& |A(\Xi_c^{0A} \to \Xi^0 K^0)|^2,  \non\\
|A(\Xi_c^{0A}\to \Sigma^+ \pi^-)|^2 &=& |A(\Xi_c^{0A} \to p
K^-)|^2,   \\
|A(\Xi_c^{+A}\to p\bar{K}^0)|^2 &=& |A(\Lambda_c^+\to \Sigma^+ K^0)|^2,
\non\\
|A(\Xi_c^{+A}\to \Xi^0 K^+)|^2 &=& |A(\Lambda_c^+ \to n \pi^+)|^2, \non\\
|A(\Xi_c^{0A}\to \Lambda^0\eta_8)|^2 &=&|A(\Xi_c^{0A} \to \Sigma^0
\pi^0)|^2,   \non
\en
for quark-mixing-suppressed modes, 
\be
|A(\Xi_c^{+A}\to \Sigma^+K^0)|^2 &=&2|A(\Xi_c^{0A} \to \Sigma^0
K^0)|^2,\non\\
|A(\Xi_c^{0A}\to \Sigma^-K^+)|^2 &=& 2|A(\Xi_c^{+A} \to \Sigma^0
K^+)|^2,   
\en
for  quark-mixing-doubly-suppressed modes, 
and relations between the squares of quark-mixing-allowed, -suppressed, and 
 quark-mixing-doubly-suppressed amplitudes:
\be
|A(\Lambda_c^+ \to p \eta_0)|^2 &=& s_1^2|A(\Lambda_c^+ \to \Sigma^+
\eta_0)|^2,   \non \\
|A(\Xi_c^{0A}\to \Xi^-K^+)|^2 &=&s^2_1|A(\Xi_c^{0A} \to \Xi^-
\pi^+)|^2,\non\\
|A(\Xi_c^{0A}\to \Sigma^+\pi^-)|^2 &=& s^2_1|A(\Xi_c^{0A} \to \Sigma^+
K^-)|^2,   \non \\
|A(\Xi_c^{+A} \to \Sigma^+ K^0)|^2 &=& s^4_1|A(\Lambda_c^+ \to p
\bar{K}^0)|^2,   \non \\
|A(\Xi_c^{+A}\to n\pi^+)|^2 &=& s^4_1|A(\Lambda_c^+ \to \Xi^0 K^+)|^2,\non\\
|A(\Xi_c^{0A} \to \Sigma^- K^+)|^2 &=& s^4_1|A(\Xi_c^{0A} \to \Xi^-
\pi^+)|^2,   \\
|A(\Xi_c^{0A}\to p\pi^-)|^2 &=&s^4_1|A(\Xi_c^{0A} \to \Sigma^+ K^-)|^2,\non\\
|A(\Lambda_c^+\to nK^+)|^2 &=&s^4_1|A(\Xi_c^{+A} \to \Xi^0 \pi^+)|^2,\non\\
|A(\Lambda_c^+\to pK^0)|^2 &=& s^4_1|A(\Xi_c^{+A} \to \Sigma^+
\bar{K}^0)|^2,   \non \\
|A(\Xi_c^+ \to p \eta_0)|^2 &=& s_1^4|A(\Lambda_c^+ \to \Sigma^+
\eta_0)|^2,   \non
\en
where $s_1=\sin\theta_1$, and $\theta_1$ is the usual quark-mixing angle.

Note that the above quark-diagram relations can also be reproduced in the 
SU(3) Hamiltonian approach of Savage and Springer (SS) [6]
\footnote{Note that the reduced matrix elements $a,~b,~c$ and $d$ introduced
in
Ref.[6] are associated with the operator $O_{\overline{15}}$, which
transforms
as a $\overline{15}$ under flavor SU(3) and is symmetric in color indices and
hence cannot induce a baryon-baryon transition. In other words, {\it 
baryon-pole diagrams are prohibited by the operator $O_{\overline{15}}$.}}
except for Eq.~(59) and the first and last relations in Eq.~(60). We 
believe that when the use of the SU(3)
Hamiltonian in which the symmetry amplitudes are tensor decomposed
is done correctly to incorporate the symmetry properties of the baryon 
wave function, the reduced matrix element $a$ defined in Ref.[6] should not
contribute and all aforementioned SU(3) quark-diagram results will be 
reproduced. 

  The relations between the SU(3) reduced matrix elements of Ref.[6] and 
the quark-diagram amplitudes are\footnote{
Using Table 2 and the relations (61), one can perform a cross check
on the SU(3) amplitudes given in Tables I-III of Ref.[6]. For example, we
find a sign error in Table III, namely the squared matrix elements for 
$\Xi_c^0\to \Lambda^0K^0$ should read ${1\over 6}|a-2b+c+2e-4f-4g|^2$.}
\be
&& a=0,~~~~~b=-{1\over 4}(\cap+\cta)-{1\over 4\sqrt{3}}(\csp+\cts),  \non\\
&& c={1\over 4}(\aa+\bap)-{1\over 2\sqrt{3}}(\csp+\cts),~~~~~
d={1\over 2\sqrt{3}}(\csp+\cts),   \non  \\
&& e={1\over 8}(\aa-\bap)+{1\over 4\sqrt{3}}(\csp-\cts),   \\
&& f=-{1\over 8}(2\coa+\cap-\cta)+{1\over 8\sqrt{3}}(\csp-\cts),   \non \\
&& g={1\over 8}(\aa+2\ba-\bap).   \non
\en
At first sight, it appears that there are six independent SU(3) parameters,
but eight different quark amplitudes. However, one may make the following
redefinition (this redefinition is not unique):
\be
\tilde{\cal A} &=& \aa-{2\over\sqrt{3}}\csp,~~~~\tilde{\cal
B}'=\bap-{2\over \sqrt{3}}\cts,~~~\tilde{\c}_S=\csp+\cts,   \non \\
\tilde{\cal C}' &=& \cap-{1\over\sqrt{3}}\csp+\coa,~~~\tilde{\cal
C}_{2}=\cta-
{1\over\sqrt{3}}\cts-\coa,
\en
so that the amplitudes for the decay modes in Table 2 can be expressed in 
terms of the six quark-diagram terms $\tilde{\cal A},~\tilde{\cal B}'$, 
$\ba,~\tilde{\cal C}',~\tilde{\cal C}_{2},~\tilde{\c}_S$. 

   Table~2.a for quark-mixing-allowed decays $B_c(\bar 3)\to B(8)+M(8)$
was also previously considered by Kohara [10]. 
In Kohara's results there are 
eight quark diagrams $a_K,~b_K,$ $c_K$,
$d_{1K}$, $d_{2K}$, $d_{3K},~d_{4K}$ and $e_K$.\footnote{In order to 
avoid notation confusion with the SU(3) parameters of SS [6], we
add a subscript $K$  to the Kohara's quark diagram amplitudes [10].}
The relations between our quark diagram amplitudes and those in [10] are
\be
&& a_K={\sqrt{6}\over 8}\aa,~~~b_K={\sqrt{6}\over 8}\bap,~~~c_K={\sqrt{6}
\over 4}\ba,   \non \\    
&& d_{1K}={\sqrt{6}\over 4}\cap-{1\over 2\sqrt{2}}\csp,~~~d_{2K}={1\over
\sqrt{2}}\csp,   \\
&& d_{3K}={\sqrt{6}\over 4}\cta-{1\over 2\sqrt{2}}\cts,~~~d_{4K}={1\over
\sqrt{2}}\cts,~~~e_K={\sqrt{6}\over 4}\coa,   \non
\en
which are obtained by comparing Table I of [10] with our Table 2a.
As emphasized before, we consider it 
conceptually clearer and in practice simpler to  work with the orthonormal
basis of quark states.

\vskip 0.3cm
\pagebreak
\noindent{\bf V. Sextet Charmed Baryon Decays}

\noindent{\bf V.a. Quark Diagram Scheme for $B_c(6) \to B(10) +M(8) $}

There are six independent quark-diagram amplitudes for
$B_c(6) \to B(10) +M(8) $. The amplitudes ${\cal B}$
and ${\cal C}_1$ are forbidden owing to the Pati-Woo theorem.
The relevant diagrams and amplitudes are exhibited in Fig. 3 and Table 3,
respectively.

   From Table 3 we obtain the following  relations, in the absence of both
SU(3)
violations and final state interactions:
\be
|A(\Omega_c^0 \to \Sigma^{*+} K^-)|^2 &=& 2|A(\Omega_c^0 \to \Sigma^{*0}
\bar{K}^0)|^2,   \non \\
|A(\Omega_c^0 \to \Sigma^{*0} \eta_0)|^2 &=& 2|A(\Omega_c^0 \to \Sigma^{*0}
\eta_8)|^2,   \non \\
|A(\Omega_c^0 \to \Sigma^{*0}\eta_0)|^2 &=& {1\over 2}s_1^2|A(\Omega_c^0 \to 
\Xi^{*0} \eta_0)|^2,    \\
|A(\Omega_c^0 \to \Sigma^{*+}\pi^-)|^2 &=& s_1^2|A(\Omega_c^0 \to \Sigma^{*+}
K^-)|^2,   \non \\
|A(\Omega_c^0 \to \Xi^{*-} K^+)|^2 &=& {1 \over 3}s_1^{2}|A(\Omega_c^0 \to
\Omega^- K^+)|^2.   \non 
\en
It is interesting to note that the $\Omega_c^0$
decays into  $\Delta^0 \bar{K}^0$ and $\Delta^+ K^-$ are prohibited
in the quark-diagram scheme
\be
|A(\Omega_c^0\to\Delta^0\bar{K}^0)|^2=0,~~~|A(\Omega_c^0\to\Delta^+K^-)|^2
=0,
\en
as the quark diagram ${\cal C}_1$ is not allowed by the Pati-Woo theorem.
Consequently, the corresponding reduced matrix element $\alpha$ makes no 
contribution.

    We note that the above quark-diagram relations except for the last 
one listed in (65) 
cannot be reproduced in the SU(3)-IR approach of SS unless the 
reduced matrix elements $\alpha$ and $\delta$ do not contribute. Therefore,
in the SU(3) limit there are only four independent quark-diagram amplitudes
or reduced matrix elements.
Relations between the quark-diagram amplitudes and the
symmetry parameters (see Eq.~(25) of Ref.[6]) 
are given by
\be
{\cal A}_S &=& \beta -2\eta,~~~~~~
{\cal B}'_S=\beta+2\eta,   \non \\
{\cal C}'_S &=& \gamma+2\lambda,~~~~~~
{\cal C}_{2S}=\gamma-2\lambda.  
\en

\noindent{\bf V.b. Quark Diagram Scheme for $B_c(6) \to B(8) +M(8) $}

We discuss in this section the decays of sextet charmed baryons into an
octet baryon and a pseudoscalar meson. 
The relevant quark diagrams and amplitudes are shown in Fig. 4 and Table 4,
respectively.

In the SU(3)-symmetry approach [6], there exist no any relations between the
decays of $\Omega_c^0 \to B(8) + M(8)$. However, from Table 4 we obtain
\be
|A(\Omega_c^0 \to n \bar{K}^0)|^2 &=& |A(\Omega_c^0 \to p K^-)|^2, \non \\
|A(\Omega_c^0 \to \Sigma^+ K^-)|^2 &=& 2|A(\Omega_c^0 \to \Sigma^0\bar{K}^0
)|^2.
\en
These relations cannot be reproduced in the SU(3) approach [6]
unless the contributions due to the SU(3) parameters
${\it a}$ and ${\it d}$ vanish. Therefore, Eq.~(67) will
provide a good test on the quark-diagram scheme.
Unfortunately, these processes are either singly or 
quark-mixing-doubly-suppressed. We do not expect that an encouraging
experimental
verification will come out soon.

    The relations between quark-diagram amplitudes and SU(3) reduced matrix
elements are found to be
\be
&& a=d=0,~~~~~
b=\,-{1\over 2\sqrt{3}}(\as+\bsp),   \non \\
&& c+l=\,{1\over 4}(\cap+\cta)-{1\over 4\sqrt{3}}(\csp+\cts),  \non \\
&& e-l=\,{1\over 2\sqrt{3}}(\csp+\cts),~~~
f=\,{1\over 8}(\cap-\cta)-{1\over 8\sqrt{3}}(\csp-\cts),   \\
&& g=\,{1\over 4\sqrt{3}}(\as-\bsp)+{1\over 4\sqrt{3}}(\csp-\cts),  \non \\
&& h=\,-{1\over 8}(\cap-\cta)+{1\over 8\sqrt{3}}(\csp-\cts)+{1\over 4}\coa,
\non \\
&& k=-{1\over 4\sqrt{3}}(\as-\bsp)-{1\over 4}\ba.  \non
\en
Therefore, there are seven independent SU(3) parameters and quark-diagram
amplitudes.

\noindent{\bf VI. Conclusions}

In this paper we have given a general and unified formulation useful for the
quark-diagram scheme for  baryons. Here we apply it to the two-body 
nonleptonic weak decays of charmed
baryons and express their decay amplitudes in terms of the quark-diagram
amplitudes.  The effects of final-state interactions and SU(3) violation 
arising in the 
horizontal $W$-loop quark diagrams 
are included in the tables. In the absence of SU(3) violation and
final-state interactions we have obtained many relations among 
various decay modes.  These relations provide a
framework to study these effects.
Some of
the relations are valid even in the presence of final-state interactions when
each
decay amplitude in the  relation contains only a single phase shift.  It will
be interesting to 
 compare all these relations with future
experimental data.

Our results are consistent with those from the SU(3)-IR scheme. In
addition,  in the quark-diagram scheme we are able to  impose the Pati-Woo
theorem
for weak decays and  obtain more specific results than those from the
SU(3)-IR scheme.

We also note that the quark-mixing-allowed decays of the antitriplet 
charmed baryon into a decuplet baryon and a pseudoscalar meson can only 
proceed through the ${\it W}$-exchange diagram.
Hence, the experimental measurement of
$\Lambda_c^+ \to \Delta^{++} K^- $ implies that the ${\it W}$-exchange
mechanism plays a significant role in charmed baryon decays.

\vskip 3.0 cm
\pagebreak
\centerline{\bf ACKNOWLEDGMENTS}

   This work was supported in part by the U.S. Department of Energy
and the National Science Council of Taiwan under Contract
No. NSC85-2112-M-001-010.
\vskip 3.0 cm
\pagebreak
%
%
%
\vskip 1.5 cm
\centerline{\bf REFERENCES}
\begin{enumerate}

\item For a review of charmed baryons, see J.G. K\"orner and H.W.
Siebert, {\sl Annu. Rev. Nucl. Part. Sci.} {\bf 41}, 511 (1991);
S.R. Klein, {\sl Int. J. Mod. Phys.} {\bf A5}, 1457 (1990); J.G. K\"orner, M.
Kr\"amer, and D. Pirjol, {\sl Prog. Part. Nucl. Phys.} {\bf 33}, 787 (1994).

\item Particle Data Group, \pr {\bf D50}, 1173 (1994).

\item J.G. K\"orner, G. Kramer, and J. Willrodt, \pl {\bf 78B}, 492 (1978);
 {\sl Z.~Phys.} {\bf C2}, 117 (1979);
B. Guberina, D. Tadi\'c, and J. Trampeti\'c, \zp {\bf C13}, 251 (1982);
F. Hussain and M.D. Scadron, {\sl Nuovo Cimento} {\bf 79A}, 248 (1984); F.
Hussain and K. Khan, {\sl ibid.} {\bf 88A}, 213 (1985);
D. Ebert and W. Kallies, {\sl Phys. Lett.} {\bf 131B}, 183
(1983); {\bf 148B}, 502(E) (1984); {\sl Yad. Fiz.} {\bf 40}, 1250 (1984);
{\sl Z. Phys.} {\bf C29}, 643 (1985); H.Y. Cheng, \zp {\bf C29}, 453 (1985); 
Yu.L. Kalinovsky, V.N. Pervushin, G.G. Takhtamyshev, and N.A. Sarikov,
{\sl Sov. J. Part. Nucl.} {\bf 19}, 47 (1988).

\item S. Pakvasa, S.F. Tuan, and S.P. Rosen, \pr {\bf D42}, 3746 (1990); M.P.
Khanna and R.C. Verma, \zp {\bf C47}, 275 (1990); G. Kaur and M.P. Khanna,
\pr {\bf D44}, 182 (1991); G. Turan and J.O. Eeg, \zp {\bf C51}, 599 (1991).

\item H.Y. Cheng and B. Tseng, \pr {\bf D46}, 1042 (1992); {\bf
D48}, 4188 (1993); Q.P. Xu and A.N. Kamal, {\sl ibid.} {\bf D46}, 270
(1992);
J.G. K\"orner and M. Kr\"amer, \zp {\bf C55}, 659 (1992); T. Uppal, R.C.
Verma,
and M.P. Khanna, \pr {\bf D49}, 3417 (1994); P. Zenczykowski, \pr {\bf D50},
410 (1994).

\item M.J. Savage and R.P. Springer, \pr {\bf D42}, 1527 (1990).

\item S.M. Sheikholeslami, M.P. Khanna, and R.C. Verma, \pr
{\bf D43}, 170 (1991); M.P. Khanna and R.C. Verma, hep-ph/9506394.

\item L.-L. Chau, in {\it Proceedings of the VPI Workshop on Weak
Interactions}, AIP Conf. Proc. No.
72, Particles and Fields, Subseries No. 23, eds. G.B. Collins, L.N. Chang,
J.R. Ficence, Blacksburg,
Virginia, December 3--6, 1980; {\sl Phys.~Rep.} {\bf 95}, 1 (1983).

\item L.-L. Chau and H.Y. Cheng, \prl {\bf 56}, 1655 (1986);
~{\sl  Phys. Rev.} {\bf D36}, 137 (1987);
~{\sl ibid.} {\bf D39}, 2788 (1989); {\sl Phys. Lett.} {\bf 222B}, 285
(1987);
{\sl Mod. Phys. Lett.} {\bf A4}, 877 (1989).

\item Y. Kohara, \pr {\bf D44}, 2799 (1991).

\item J.C. Pati and C.H. Woo, \pr {\bf D3}, 2920 (1971);
K. Miura and T. Minamikawa {\sl Prog. Theor. Phys.} {\bf 38}, 954 (1967);
J.G.
K\"orner, \np {\bf B25}, 282 (1970); {\sl Z. Phys.} {\bf C43},
165 (1989).

\item See, e.g., F.E. Close, {\it An Introduction to Quarks and 
Partons} (Academic Press, 1979).

\end{enumerate}
\pagebreak

%
%
\vskip 1.5 cm
\centerline{\bf FIGURE CAPTIONS}
\vskip 0.4 cm
\begin{description}
\item[Fig.~1.] Quark diagrams for the decay ~$B_c(\bar 3) \to B(10) +M(8) $.

\item[Fig.~2.] Quark diagrams for the decay ~$B_c(\bar 3) \to B(8) +M(8) $.

\item[Fig.~3.] Quark diagrams for the decay ~$B_c(6) \to B(10) +M(8) $.

\item[Fig.~4.] Quark diagrams for the decay ~$B_c(6) \to B(8) +M(8) $.
\vskip 1.5 cm
\centerline{\bf TABLE CAPTIONS}
\vskip 0.4 cm
\item[Table 1a.]  Quark-diagram amplitudes for the quark-mixing-allowed 
decays of $B_c(\bar 3) \to B(10) +M(8)$.  In Tables 1-3, SU(3) symmetry
breaking is parametrized only in terms of the $\e$ type of quark diagrams. 

\item[Table 1b.]  Quark-diagram amplitudes for the quark-mixing-suppressed
decays of $B_c(\bar 3) \to B(10) +M(8) $.

\item[Table~1c.]  Quark-diagram amplitudes for the 
quark-mixing-doubly-suppressed
decays of $B_c(\bar 3) \to B(10) +M(8) $.

\item[Table~2a.]  Quark-diagram amplitudes for the quark-mixing-allowed 
decays of $B_c(\bar 3) \to B(8) +M(8) $.

\item[Table~2b.]  Quark-diagram amplitudes for the quark-mixing-suppressed
decays of $B_c(\bar 3) \to B(8) +M(8) $.

\item[Table~2c.]  Quark-diagram amplitudes for the 
quark-mixing-doubly-suppressed
decays of $B_c(\bar 3) \to B(8) +M(8) $.

\item[Table~3.]  Quark-diagram amplitudes for $B_c(6) \to B(10) +M(8) $.

\item[Table~4.]  Quark-diagram amplitudes for $B_c(6) \to B(8) +M(8) $.

\end{description}


\pagebreak

\begin{figure}[h]
\hskip 0cm
\hbox{
\epsfxsize=16cm
\epsfysize=20cm
\epsffile{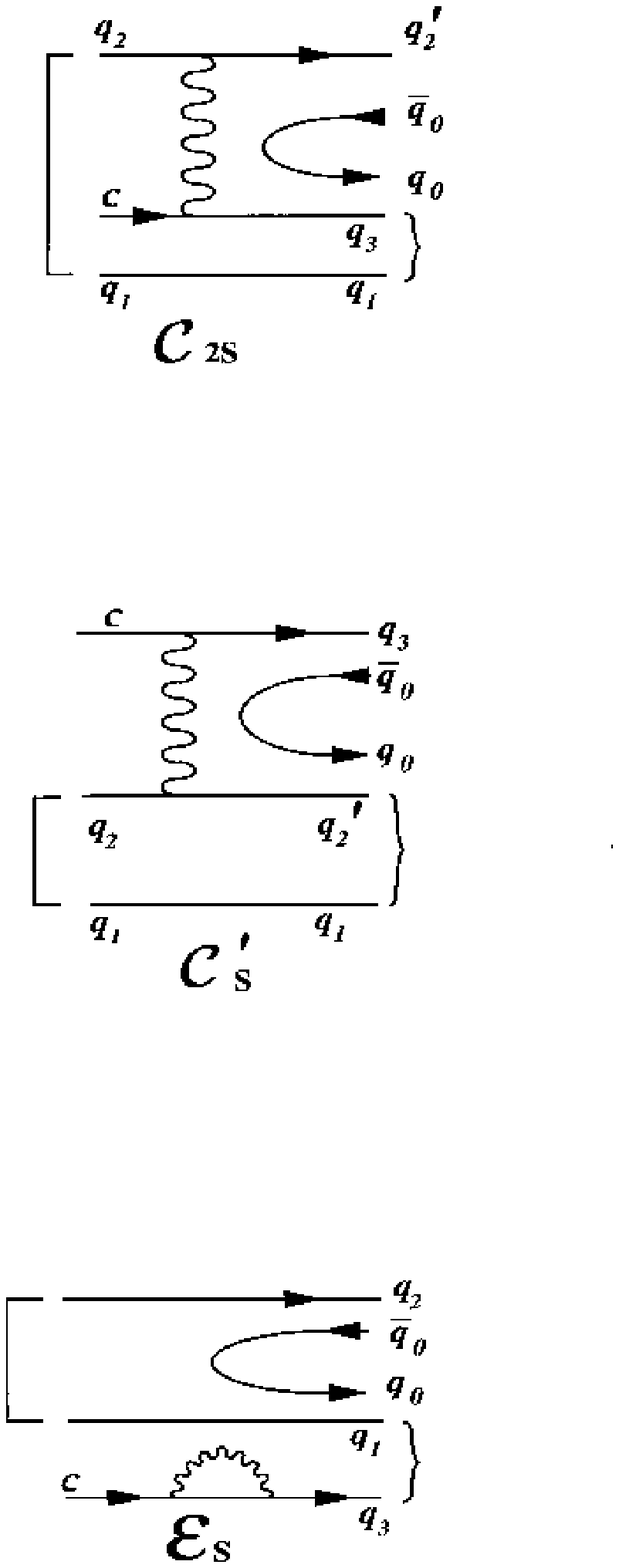}} 
\end{figure}

\newpage

\begin{figure}[h]
\hskip 1cm
\hbox{
\epsfxsize=16cm
\epsfysize=20cm
\epsffile{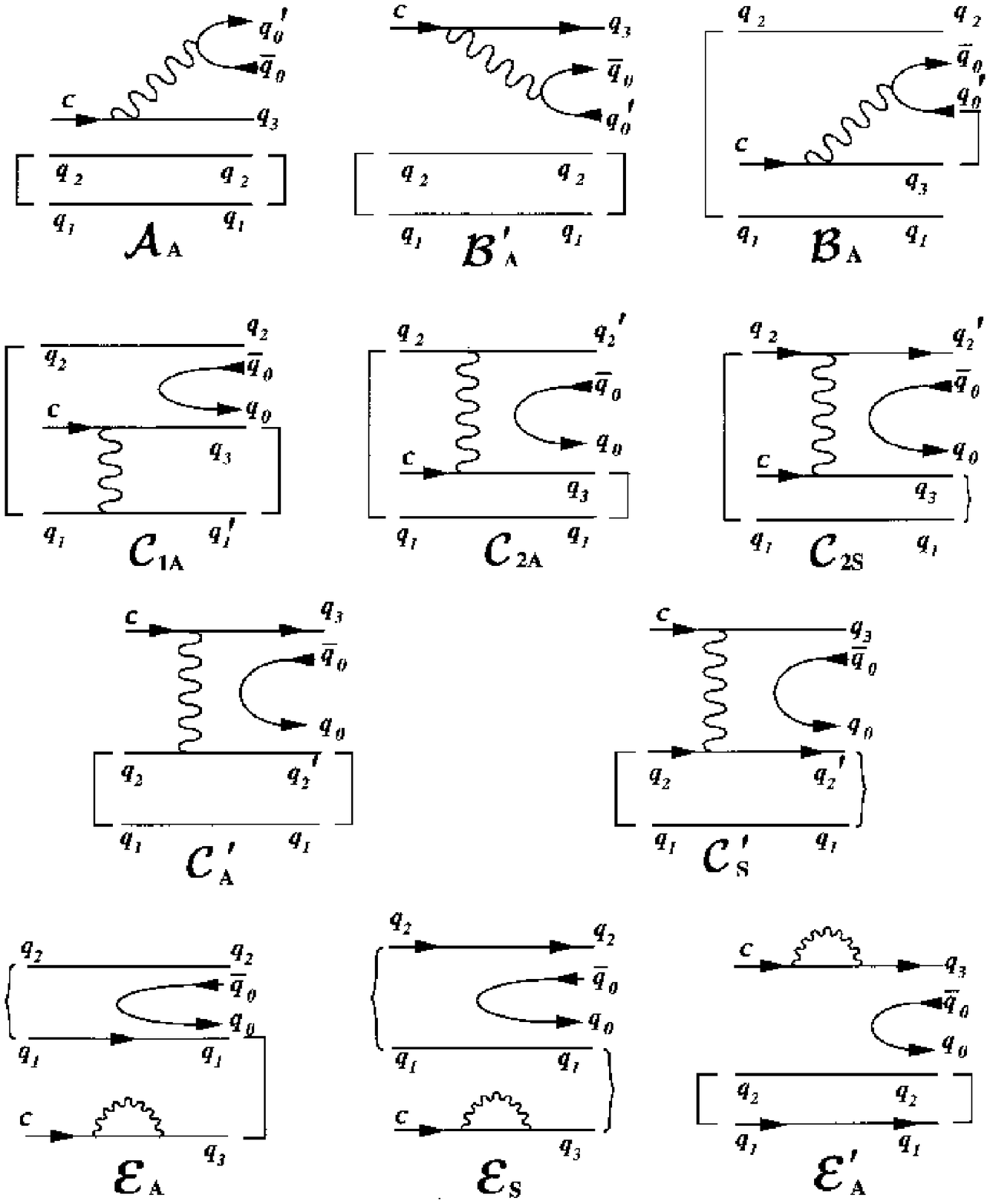}} 
\end{figure}

\newpage

\begin{figure}[h]
\hskip 1cm
\hbox{
\epsfxsize=16cm
\epsfysize=20cm
\epsffile{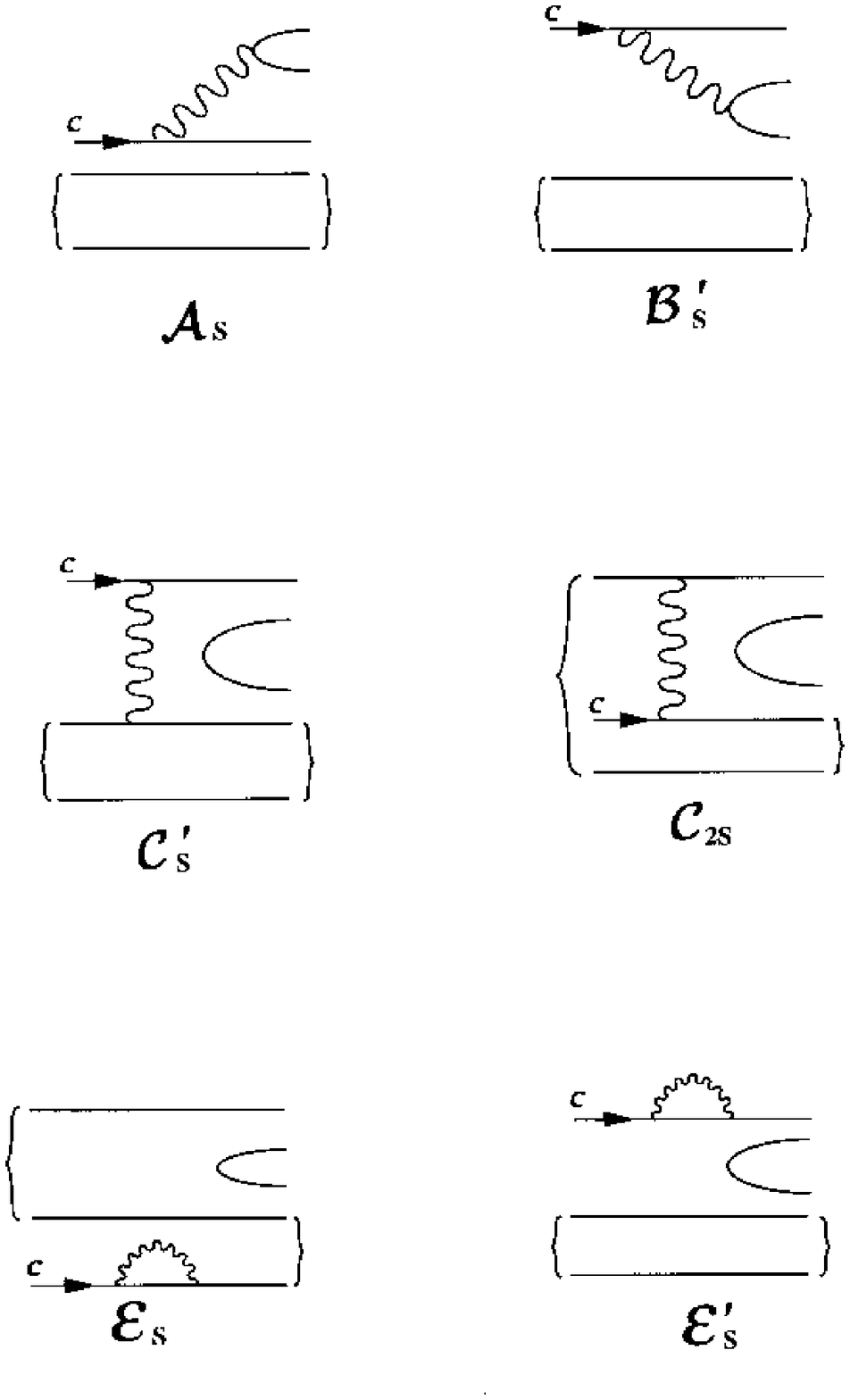}} 
\end{figure}

\newpage

\begin{figure}[h]
\hskip 1cm
\hbox{
\epsfxsize=16cm
\epsfysize=20cm
\epsffile{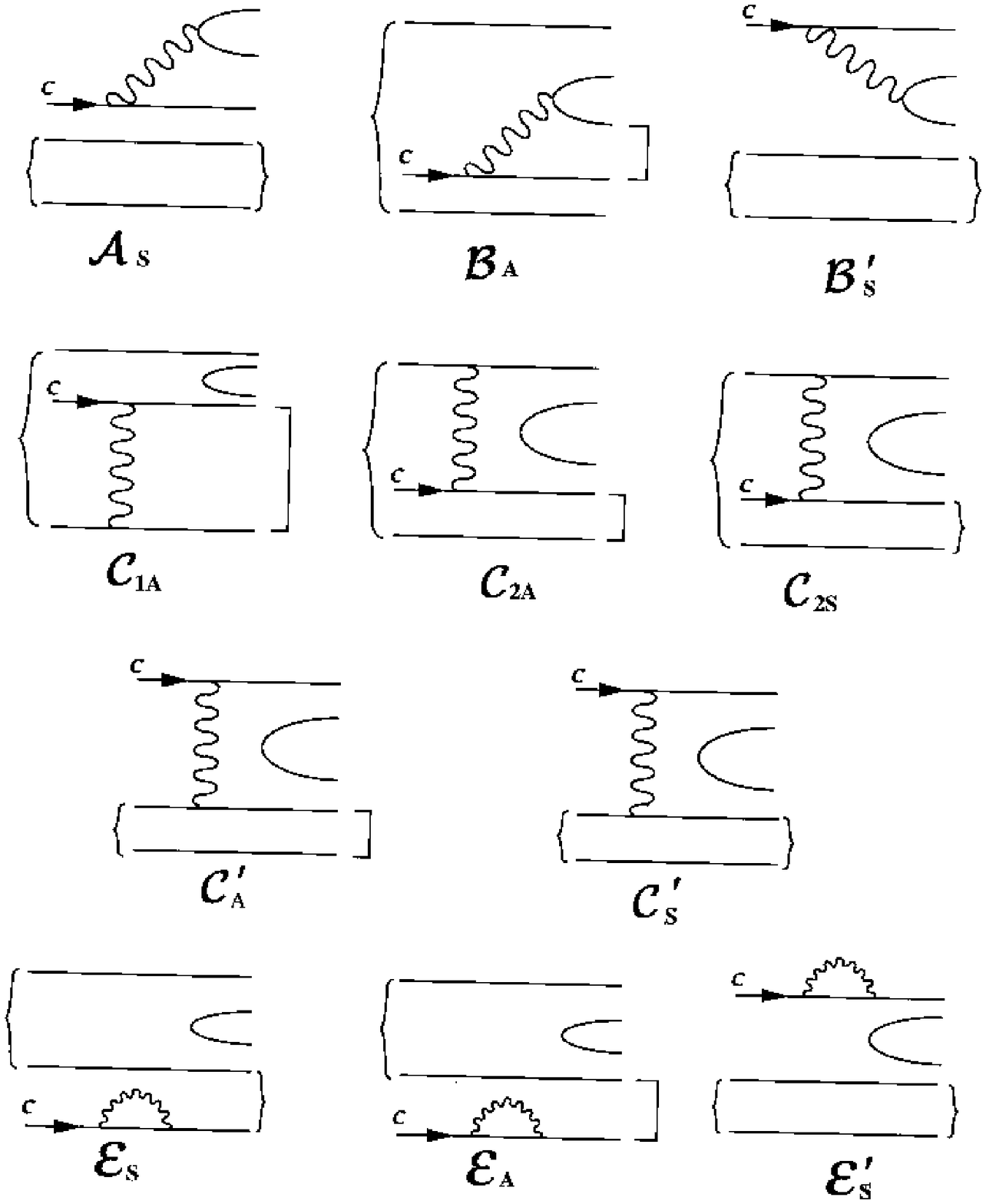}} 
\end{figure}

\newpage

\end{document}